# Vesicle-surface-templated catalytic polymers drive differential growth in synthetic minimal cell variants.

Minoru Kurisu[a,b,*], Taro Suzuki[a], Ryosuke Katayama[a], Kazuki Maruyama[a], Daisuke Unabara[c2], Tasuku Hamaguchi[c], Koji Yonekura[c,d], Peter Walde[e], and Masayuki Imai[a,]

[a] Department of Physics, Graduate School of Science, Tohoku University, 6-3 Aramaki, Aoba, Sendai 980-8578, Japan.

[b] Department of Physics, Faculty of Science, Kyushu University, 744 Motooka, Nishi-ku, Fukuoka 819-0395, Japan.

[c] Institute of Multidisciplinary Research for Advanced Materials, Tohoku University, 2-1-1 Katahira, Aoba-ku, Sendai 980-8577, Japan.

[d] RIKEN SPring-8 Center, 1-1-1 Kouto, Sayo, Hyogo 679-5148, Japan.

[e] Department of Materials, ETH Zürich, Leopold Ruzicka-Weg 4, CH-8093 Zürich, Switzerland.

*Corresponding author. E-mail: kurisu.minoru.877@m.kyushu-u.ac.jp

[1] orcid.org/0000-0001-5864-320X

[2] orcid.org/0000-0001-5347-7257

[3] orcid.org/ 0000-0003-0876-7700

[4] orcid.org/0000-0001-5520-4391

[5] orcid.org/0000-0002-0827-0545

[6] orcid.org/0000-0002-1400-7794

## Abstract:

Understanding how life-like behaviors can emerge from simple molecular assemblies and primitive compartments remains a central challenge in origins-of-life research. Synthetic minimal cells provide a bottom-up platform for investigating, from scratch, the minimal physicochemical principles underlying compartment growth, reproduction, and evolution. Previously, we developed a vesicle/polymer-based compartment system in which the vesicle membranes template the formation of a catalytic polymer. This polymer promotes selective incorporation of amphiphiles into the vesicle membrane, driving vesicle growth while maintaining the compositional identity and enabling spontaneous deformation and division over several generations. Here, we report about experiments in which we advanced this system beyond reproduction by systematically constructing eight synthetic minimal cell variants from combinations of two template vesicles, two catalytic polymers, and two supplied amphiphiles. The variants exhibited distinct, composition-dependent vesicle growth responses, ranging from pronounced growth to suppressed growth or vesicle shrinkage. These growth responses were described by the Hill kinetics and characterized by three parameters, revealing a multi-dimensional fitness landscape shaped by environmental conditions, in which the relative advantage of each variant depends on both composition and amphiphile availability. This framework links molecular recognition, compositional inheritance, and differential growth, providing a physicochemical route toward evolvable synthetic minimal cells.

## Introduction

Understanding how living systems emerged from non-living molecular assemblies remains a fundamental challenge in contemporary science[1–6]. Bottom-up studies on protocells (hypothetical precursor compartment systems of the first cells), artificial cells, and synthetic minimal cells provide a powerful strategy to address this problem by reconstructing essential cellular functions, such as metabolism, reproduction, and evolution, in simple vesicle-based systems.

Two major streams have emerged in this field. The first aims to reconstruct cellular functions within artificial compartments by leveraging modules from extant life, particularly gene expression systems such as the PURE system and the TX-TL (transcription-translation) platforms[7–10]. This class of system is often referred to as biological-module-based minimal cells. As gene expression can rapidly generate defined sets of proteins, this approach enables efficient implementation of sophisticated functions. However, by presupposing the machinery responsible for gene expression, heredity, and evolution, this framework cannot directly address how such mechanisms originally emerged from prebiotic molecular assemblies.

The second stream comprises synthetic minimal cells, which seek to capture minimal features associated with living systems from simple molecular components and physicochemical processes, without relying on the complex molecular machineries developed in extant life. By drastically simplifying the system components, this approach is well suited for identifying the minimal physicochemical principles required for life-like behaviors. However, constructing such systems de novo remains inherently challenging. To date, most studies of synthetic minimal cells have focused on vesicle formation, growth, and division driven by amphiphile synthesis[5,11–15]. In contrast, experimental platforms that explicitly couple the compartment dynamics to heredity and evolutionary processes remain limited.

Recent studies have begun to show that molecular recognition can induce selective vesicle growth. For example, Baba and co-workers[16] found that, in systems where fatty acid vesicles coexist with short peptides, the peptides with specific amino acid sequences promote vesicle growth in response to the external supply of fatty acids. This finding indicates that specific interactions between fatty acids and peptides can generate differential vesicle growth, potentially leading to competition and selection. Thus, in environments with diverse peptide sequences, protocellular systems incorporating sequences that confer higher vesicle growth rates (*i.e.*, corresponding to higher fitness) may be preferentially maintained, providing an experimental basis for primitive forms of heredity and evolution.

For exploring how selective compartment growth can be coupled to catalytic polymers, we have developed a vesicle-polymer platform in the field of synthetic minimal cells that links molecular recognition to vesicle growth[17–19]. In this system, the vesicle membrane acts as a template for polymer synthesis through specific interactions between sulfonated amphiphiles (**Fig.1a**; sodium bis(2-ethylhexyl) sulfosuccinate, AOT) and protonated secondary amine groups in the polymer (**Fig.1b**; polyaniline in its emeraldine salt form, PANI-ES formed from aniline). At the same time, these interactions guide the selective incorporation of amphiphiles through the catalytic polymer, resulting in vesicle growth. On the AOT vesicle surface, the vesicle-surface-assisted template polymerization of aniline is directed to suppress branched side products and favor the linear head-to-tail nitrogen-carbon (NC) linkage of aniline via specific vesicle-to-polymer interactions[20–23]. The resulting PANI-ES preferentially promotes incorporation of externally supplied AOT into the vesicle membrane, driving vesicle growth while preserving the identity of the template vesicle. In other words, the template vesicle guides polymer synthesis, and the synthesized polymer, in turn, enhances vesicle growth, establishing a mutually catalytic relationship that provides potential basis for mechanisms of heredity and evolution. Importantly, by introducing an inverse-cone-shaped lipid (*e.g.*, cholesterol or phosphatidylethanolamine, PE) as a second component for the template AOT vesicle, this vesicle-polymer system exhibits spontaneous deformation and division over four to five generations, coupled with membrane growth, *i.e.*, reproduction of a synthetic minimal cell[18,24].

Here, we advance this platform to systematically examine whether synthetic minimal cells can be extended toward heritable, competitive, and evolution-like behaviors. Specifically, we ask whether (i) multiple types of catalytic polymers can be formed on template vesicle surfaces, (ii) the types and rates of amphiphile incorporation depend on the catalytic polymer, and (iii) such differences can generate competitive advantages among minimal cell variants. To address these questions, we introduce sodium dodecylbenzenesulfonate (SDBS; **Fig.1a**) as an additional amphiphile and polypyrrole in its bipolaronic/polaronic form (PPy; **Fig.1b**) as an additional catalytic polymer, both known to form vesicle-assisted template-polymerization systems analogous to AOT/PANI-ES[25–28]. Catalytic polymers (PANI-ES or PPy) are synthesized on two types of template vesicles composed of either AOT or SDBS+decanoic acid (DA; 1/1 molar ratio). After polymer formation, micellar solutions of AOT or SDBS are supplied to the resulting minimal cell systems, and vesicle sizes before and after amphiphile addition are measured by dynamic light scattering (DLS). By quantitatively comparing the growth rates of eight minimal cell variants, comprising the combinations of two template vesicles, two catalytic polymers, and two supplied amphiphiles, we investigate whether differences in growth rate can provide competitive advantages, thereby establishing a route toward synthetic minimal cells capable of heredity- and evolution-like dynamics.

## Results

### Vesicle-surface template polymerization generates multiple synthetic minimal-cell variants

In the presence of small (sonicated) unilamellar vesicles (SUVs) composed of sulfonated amphiphiles, the oxidative enzymatic polymerization of aniline or pyrrole is known to yield the electroconductive forms of polyaniline (the emeraldine salt form; PANI-ES) or polypyrrole (PPy), as long as ideal reaction conditions are applied (**Fig. 1b**), as previously reported[17,20,28,29]. In this study, we used sodium bis(2-ethylhexyl) sulfosuccinate (AOT) vesicles and binary sodium dodecylbenzenesulfonate (SDBS) + decanoic acid (DA) (1/1, molar ratio) vesicles as template vesicles for the enzymatic synthesis of PANI-ES and PPy (**Fig. 1b**). While AOT molecules form vesicles in 20 mM $NaH_2PO_4$ aqueous solution, SDBS molecules do not form single-component vesicles. However, if prepared with equimolar amounts of DA molecules, SDBS and DA molecules form binary SDBS+DA (1/1) vesicles in 20 mM $NaH_2PO_4$ aqueous solution. Both types of vesicles, from AOT or SDBS+DA (1/1), were used as templates for the enzymatic polymerization of aniline or pyrrole. The vesicle dispersions were first prepared by the gentle hydration method (see **Methods**), which resulted in vesicles with high size polydispersity, ranging in diameters from sub-micrometers to tens of micrometers. Then, the vesicle dispersions were treated with ultrasound to reduce and homogenize the vesicle diameter, yielding average hydrodynamic vesicle sizes of about 90 nm, as determined by dynamic light scattering (DLS) measurements. Thus, relatively uniform SUVs composed of AOT or SDBS+DA (1/1) were obtained, which were then used as template vesicles for the polymerization reactions. In the presence of these template SUVs, the enzymatic polymerizations of aniline or pyrrole were carried out at room temperature ($T$ ~ 25˚C) with a reaction time of $t$ = 24 h, following the previously reported procedures (see **Methods**). The sulfonated vesicle surfaces provide reaction fields for the synthesis of PANI-ES or PPy and “guide” the polymerization reactions in a positive way, *i.e.*, the vesicles (i) act as counter ions, (ii) prevent the formation of branched polymer chains, (iii) prevent the formation of over-oxidized or hydrolyzed products to yield the conductive forms of the polymers in a desired oxidation state, and (iv) keep the obtained polymers homogenously dispersed within the vesicle dispersions[20,22,23,28,30].

In the present work, the successful enzymatic synthesis of PANI-ES and PPy was first confirmed spectrophotometrically for the specific reaction setting used. The polymerization of aniline with horseradish peroxidase (HRP) as catalyst was triggered by the addition of $H_2O_2$. Subsequent to the initiation of the reaction with $H_2O_2$, the initially colorless reaction mixtures turned to blue within several seconds and then relatively rapidly to dark green. The UV/Vis/NIR absorption spectra of the reaction mixtures obtained with AOT SUVs or binary SDBS+DA (1/1) SUVs as templates are shown as red bold lines in **Fig. 2a** and **b**, respectively. The spectra indicate the formation of PANI-ES from aniline with characteristic absorption maxima at $\lambda$ ~ 300 nm, 420 nm, and broad absorption of high intensity around ~1,000 nm, which previously were assigned to $\pi \rightarrow \pi^*$, polaron $\rightarrow \pi^*$, and $\pi \rightarrow$ polaron transitions[31–34]. Furthermore, low absorption at $\lambda$ ~ 500 nm indicates a low content of chain branching structures[35] and/or a low extent of phenazine formation[22,23], both of which are undesirable side products. In the case of the synthesis of PPy from pyrrole with *Trametes versicolor* laccase (TvL) and dissolved $O_2$, the color of the reaction mixtures slowly turned from colorless to dark green, the color intensity steadily increasing over several hours after TvL addition, indication a much slower rate of polymerization than in the case of the polymerization of aniline with HRP/$H_2O_2$. The UV/Vis/NIR absorption spectra of the polypyrrole reaction mixtures obtained with AOT SUVs or binary SDBS+DA (1/1) SUVs are shown as blue thin lines in **Fig. 2a** and **b**, respectively. Like the spectra of the aniline reaction mixtures, the spectra of the pyrrole reaction mixtures showed a characteristic absorption maximum at $\lambda$ ~ 450 nm and a broad band around 1,000 nm, which are indicative for the presence of the conductive form of polypyrrole (PPy)[36–39]. Comparing the absorption spectra of the obtained polyaniline and polypyrrole products for the conditions used, it is evident that the absorbance of PPy was lower than the absorbance of PANI-ES. This is mainly due to differences in the chromophores and the reaction progress (reaction yield); the TvL/$O_2$-catalyzed polymerization of pyrrole is slower than the HRP/$H_2O_2$-catalyzed polymerization of aniline.

**Figure 2c** shows Raman spectra of the reaction mixture initially containing aniline and HRP/$H_2O_2$ (red bold line) and of the reaction mixture initially consisting of pyrrole and TvL/$O_2$ (blue thin line), both with AOT SUVs as templates. Several Raman bands support the UV/Vis/NIR absorption measurements in the sense that there are clear indications that the reactions yield the conductive form of PANI or PPy, respectively. For the aniline reaction mixture, the strong band at 1345 $cm^{-1}$ is one of the prominent features of the protonated conductive emeraldine salt form of polyaniline (PANI-ES), which is attributed to $C–N^{\cdot+}$ stretching with delocalized polarons (radical cation)[41–44]. The band at 1587 $cm^{-1}$ was previously assigned to C=C and C–C stretching of quinoid and semiquinonoid rings with radical cations, which is also indicative for PANI-ES formation[40–42]. The band at 608 $cm^{-1}$ is associated with in-plane ring deformations of the PANI-ES polaronic structure, combined with the deformation of the sulfonate group of the template amphiphile (AOT)[41,42]. For the pyrrole reaction mixture, bands at 939 and 983 cm$^{-1}$ are assigned to ring deformations associated with bipolaron (dication) and polaron (radical cation) units of conductive polypyrrole (PPy), respectively[45–47]. The bands at 1052 and 1081 cm$^{-1}$ are assigned to the C–H in-plane deformation with radical cation and dication, respectively, which is also indicative for the formation of conductive PPy[45–47]. The bands between 1550 and 1650 cm$^{-1}$, which have small shoulders and are difficult to resolve into individual bands, are assumed to consist of several overlapping C=C stretching bands from the neutral, radical cation, and dication forms of polypyrrole[45,46,48]. The characteristic Raman bands of PANI-ES and PPy described above for the reactions run in the presence of AOT SUVs as templates were also observed in the reaction mixture run in the presence of binary SDBS+DA (1/1) SUVs as templates (**Fig. 2d**).

All in all, we conclude that two types of polymers (PANI-ES and bipolaronic/polaronic PPy) are successfully formed on two types of sulfonated vesicle surfaces composed of either AOT or SDBS+DA (1/1)) in the reaction mixtures prepared. Thus, the vesicle-polymer composition space of our previously developed synthetic minimal cell, originally based on the AOT/PANI-ES system[17,18], is substantially expanded to include SDBS+DA (1/1) vesicle and PPy. Importantly, these newly established vesicle-polymer variants retain the ability to grow upon the external supply of AOT or SDBS micelles, as demonstrated by DLS and cryo-TEM measurements, as described below.

The average sizes of AOT SUVs and binary SDBS+DA (1/1) SUVs were determined by DLS measurements (as mean hydrodynamic vesicle diameters), and vesicle morphologies were observed by cryo-TEM (**Fig.2e-g**). The mean hydrodynamic diameters of AOT SUVs and SDBS+DA SUVs before starting the polymerization reactions (*i.e.*, 24 h after the preparation of the SUV dispersions by ultrasonication, see **Methods**) were approximately 90 nm. Cryo-TEM images of the AOT SUV dispersion before the reaction as a template showed that the vesicles were rather monodisperse and spherical, with a high degree of unilamellarity (**Fig. 2e**). However, after enzymatic PANI-ES or PPy synthesis, the sizes of AOT SUVs or binary SDBS+DA (1/1) SUVs were found to decrease (**Fig.2f** for AOT SUVs after PANI-ES synthesis). According to DLS measurements, the mean hydrodynamic diameters of AOT SUVs and binary SDBS+DA (1/1) SUVs after the synthesis of PANI-ES or PPy, respectively, were approximately 45 ± 10 nm. To these reaction mixtures containing SUVs and enzymatically formed PANI-ES or PPy, we additionally supplied AOT or SDBS molecules and monitored the size changes of AOT or SDBS + DA (1/1) SUVs before and after supplying amphiphiles using DLS, as quantitatively described in the following section. One example of such size monitoring is shown in **Fig.2h.** The slope of the intermediate scattering function of an AOT SUVs/PANI-ES reaction mixture before supplying AOT molecules (black filled circles) was dramatically decreased after supplying AOT molecules (red filled squares), indicating an increase of the mean hydrodynamic vesicle diameters. The corresponding cryo-TEM images of the SUVs before and after supplying AOT molecules are shown as **Fig.2f** and **g**, respectively, indicating that the AOT SUVs increased their mean size by supplying AOT molecules, while keeping spherical morphology.

## Catalytic polymer identity determines amphiphile-dependent vesicle growth

Sulfonated amphiphiles (AOT or SDBS) were supplied at various concentrations as micellar solutions to each reaction mixture containing one type of vesicle-polymer minimal-cell variant (*i.e.*, template SUVs and enzymatically formed PANI-ES or PPy). A shortened notation of these types of experiments is used in **Figs. 2–5**, for example "AOT(v)/PANI/AOT(m)", indicating the initial presence of AOT template vesicles ("v"), the enzymatic formation of PANI-ES, and the subsequent addition of AOT micelles ("m").

The average size of the vesicles after supplying additional amphiphiles was determined by DLS as hydrodynamic diameter. The vesicle growth experiments were carried out for eight different variants. These eight variants are the possible combinations of the two types of template SUVs (made from AOT or binary SDBS+DA (1/1)), two types of enzymatic polymerization reactions (aniline/HRP/$H_2O_2$ or pyrrole/TvL/$O_2$), and two types of added sulfonated amphiphiles (AOT or SDBS). As described in **Methods**, in each micelle addition experiment, 500 µL in total of a micellar solution (3–20 mM amphiphile) and 500 µL in total of a 40 mM $NaH_2PO_4$ solution were added within two minutes under gentle mixing into 1.0 mL reaction mixtures containing 3.0 mM amphiphiles (SUVs), enzymatically obtained PANI-ES or PPy, remaining monomers (aniline or pyrrole), and enzymes in 20 mM $NaH_2PO_4$ solution.

DLS measurements carried out 10 min after completing the addition of the AOT or SDBS micellar solutions demonstrated that the average size of the template SUVs with formed PANI-ES or PPy increased for some of the variants. The average vesicle growth estimated as relative membrane area, as a function of supplied amphiphile concentration, is shown in **Fig.3a–h**. The red and blue filled circles represent the vesicle growth observed using the typical reaction conditions for the enzymatic PANI-ES and PPy synthesis (see **Methods**). For a direct comparison of growth across the eight variants, the relative membrane area observed at 1.5 mM and 3.0 mM of supplied amphiphiles is shown in **Fig.3i** and **j**, respectively; the red and blue bars again represent vesicle growth in the presence of PANI-ES and PPy, respectively, and the labels a–h corresponds to the data shown in **Fig.3a-** **h**. It should be noted that the polymer-free control variants in the reaction mixtures (**Supplementary Figs.S1,S2**) did not exhibit the same growth enhancement as observed in **Fig.3a**-**h**. For three variants in **Fig.3** (**c, f,** and **g**), the supplied SDBS micelles did not result in vesicle growth, and two of them (**f** and **g**) resulted in distinct decrease of vesicle size, *i.e.*, the relative membrane area became less than 1.0, possibly due to a general SDBS surfactant effect causing the partial solubilization of the SUVs. For the other five variants, a sigmoidal membrane area growth occurred.

To gain further insight into the vesicle growth curves, additional growth experiments were carried out, varying the amounts of formed PANI-ES or PPy (shown in gray filled squares in **Fig.3a-h**). For the variants containing PANI-ES, the amount of the reaction trigger $H_2O_2$ was either increased by 50% (gray filled squares in **Fig.3a,d**) or decreased by 50% (gray filled squares in **Fig.3b**) with respect to the typical condition of 2.25 mM $H_2O_2$. For the variants containing PPy, the reaction time was increased to 2 days (gray filled squares in **Fig.3e,h**), as compared to the typical condition of 1 day. The increase or decrease of the amount of PANI or PPy resulted in the increase or decrease of vesicle growth rate, respectively, while retaining the overall sigmoidal profile but with a changed magnitude.

## Hill-type kinetics reveal variant-specific growth responses

The sigmoidal membrane growth observed with the original polymer amounts (red and blue filled circles in **Fig. 3a-h**) and with increased or decreased polymer amounts (gray filled squares in **Fig. 3a-h**) were fitted using the Hill equation, which often is applied to evaluate enzymatic reactions that involve a cooperative substrate-substrate effect at the reaction site of enzyme[49,50]. Here, we employ the Hill equation by considering the supplied amphiphiles as "substrates" and the polymers as "amphiphile-binding enzymes" for fitting the vesicle growth curves:

$$\Delta a([A]) = V_{\max}\frac{[A]^n}{K_{\mathrm{A}}^n + [A]^n} = \frac{V_{\max}}{1 + \left(\frac{K_{\mathrm{A}}}{[A]}\right)^n}, \tag{1}$$

where $\Delta a$ is the relative vesicle membrane increase (*i.e.*, the relative vesicle growth rate), $[A]$ is the concentration of supplied amphiphile, $V_{\max}$ is the maximal (*i.e.*, saturated) membrane growth rate, $K_{\mathrm{A}}$ is related to the polymer-amphiphile dissociation constant and is identical with the amphiphile concentration at which the reaction rate is half the value of $V_{\max}$, and $n$ is the Hill coefficient that is an index of the amphiphile-amphiphile cooperativity at the amphiphile binding site of the polymer.

In light of the general reaction scheme of the Hill equation[49], the vesicle growth promoted by PANI-ES or PPy with the addition of AOT or SDBS micellar solution in this study is schematically suggested as in **Fig.3k,l**. The special case of $n = 1$ in Eq. 1 (**Fig.3k**) means that the Hill equation is reduced to the Michaelis-Menten model (or the Langmuir-Hinshelwood model of a surface adsorption reaction; mathematically equivalent): when an amphiphile binds to a binding site of the catalytic polymer on the vesicle surface, each binding event may occur independently. Here, the first binding between an amphiphile and a binding site forming an intermediate complex state, $PA_1$ (**Fig.3k**), can be interpreted as not affecting the binding affinity between proximate binding sites and amphiphiles (*i.e.*, no cooperativity in binding), and the initially bound amphiphile would be solely incorporated into the template vesicle membrane. On the other hand, the case of $n > 1$ (**Fig.3l**) may reflect positive cooperativity in amphiphile binding, whereby the initial formation of $PA_1$ promotes the subsequent binding of additional amphiphiles (*i.e.*, cooperativity in binding), leading to the formation of $PA_{n'}$. Larger values of $n$ are indicative of higher cooperativity in the amphiphile binding, resulting in a sigmoidal response in vesicle growth rates with a steeper rise.

The vesicle growth observed with the original polymer amounts (red and blue filled circles in **Fig.3a-h**) and with increased or decreased polymer amounts (gray filled squares in **Fig.3a-h**) were consistently fitted by the Hill equation (Eq. 1), while the detailed features in growth curves are composition-dependent. The obtained values of the fitting parameters for each growth curve are listed in **Table 1**. Here, large values of $V_{\max}$ indicate high saturated growth rate after the growth curves reaching the plateau. Large values of $n$ indicate steep rise in the sigmoidal growth curves. Large values of $K_{\mathrm{A}}$ indicate high concentrations of supplied amphiphiles are required to approach $V_{\max}$, *i.e.*, the rise in vesicle growth curve shifts horizontally towards higher concentrations of supplied amphiphile. Notably, the value of the Hill coefficient was consistently obtained as $n > 2.5$ for the observed vesicle growth, implying that the supplied amphiphiles were incorporated in a “cooperative way” (**Fig.3l**) into the SUVs via the “catalytic” polymers (PANI-ES or PPy) into the SUVs. Importantly, each variant is characterized by multiple variant-specific parameter values rather than a single growth rate. Consequently, variant fitness depends on environmental conditions, such as the type and concentration of supplied amphiphiles.

For the variants resulting in enhanced vesicle growth, when the amounts of PANI-ES or PPy were increased (condition a,d,e,h; gray filled squares in **Fig.3a-h** and bracketed values in **Table 1**)) or decreased (condition b), the values of maximal growth rate $V_{\max}$ were consistently increased or decreased, respectively. This is consistent with the fact that $V_{\max}$ in the Hill equation is intrinsically proportional to enzyme concentration. Interestingly, the index value $n$ describing the substrate-substrate cooperativity seems to be conserved with respect to changes of the polymer amounts (**Table 1**; see the bracketed and unbracketed values of $n$). This consistency in the Hill coefficient $n$ implies that the molecular mechanism for incorporating the supplied amphiphiles is unique for each variant and maintained within the range of polymer amount used.

In addition to the changes of the mean vesicle size, the changes of the size distribution within the vesicle dispersions were obtained from the CONTIN analysis (inverse Laplace transform) of the intermediate scattering functions, as shown in **Fig.4a**-**e**. The vesicle size distribution after growth showed a distinct difference between the mixtures containing PANI-ES (red bold lines in **Fig. 4a-c**) and the mixtures containing PPy (blue bold lines in **Fig. 4d-e**). The template SUVs in the presence of PANI-ES grew by retaining their narrow size distribution (**Fig.4a-c**). In the case of PPy, the size distribution of the template SUVs was initially narrow, as in the case of PANI-ES, but the distribution broadened significantly upon the supply of the amphiphiles (**Fig.4d-e**). These contrasting size distributions after growth imply that the growth behavior is strongly dependent on the type of catalytic polymer: the variants containing PANI-ES exhibit relatively homogeneous growth across the vesicle population, whereas PPy-containing variants exhibit more heterogeneous growth.

To investigate whether the saturation of vesicle growth (**Fig.3a-h**) originates from saturated amphiphile binding to PANI-ES or PPy, we examined how the concentration of supplied amphiphiles affects the time required for vesicle growth to reach an equilibrium state. Focusing on the AOT(v)/PANI/AOT(m) variant (**Fig.3a**), the mean hydrodynamic radius of the template SUVs was measured by DLS at 10 min, 2 h, and 10 h after completion of micelle addition (the same procedure as in the growth experiments of **Fig.3a,** except for the waiting time). In **Fig.4f**, the black filled circles correspond to a low concentration of supplied amphiphile (1.5 mM): although the membrane area approximately doubled within the initial 10 min, no further growth was observed thereafter, indicating that the incorporation of added amphiphiles to the vesicle membranes was completed within 10 min. By contrast, at higher concentrations of supplied amphiphiles 3.0 mM (green filled squares) and 4.5 mM (red filled diamonds), the vesicle growth continued beyond 10 min, showing that the incorporation of the added amphiphiles was not yet completed within 10 min, reaching higher extents of amphiphile incorporation. These results are consistent with the membrane growth observed in **Fig.3a**, *i.e.*, the 1.5 mM supply of amphiphiles was insufficient to saturate the vesicle growth rate, while the supply of 3.0 mM or 4.5 mM amphiphiles resulted in a growth rate saturation at clearly higher levels than in the case of 1.5 mM. Interestingly, the narrow size distribution of the SUVs was maintained during and/or after vesicle growth up to as long as 10 h after completing the supply of micellar amphiphiles (**Fig.4g**).

Even after the AOT(v)/PANI/AOT(m) variant had experienced the supply of AOT micelles and had grown up to the equilibrium size, a following second supply of amphiphiles was found to trigger a further increase in SUV size. After withdrawing portions of an AOT SUV/PANI-ES mixture that already experienced vesicle growth by supplying 1.5 mM AOT micelles, 1.5 mM additional AOT micelles were again supplied 2 h and 10 h after the first supply, respectively, and then 10 min later the mean hydrodynamic radii were measured by DLS (red filled diamonds in **Fig.4h**). Similarly to the first-time vesicle growth, the AOT SUVs in both reaction mixtures (*i.e.*, the 2 h and 10 h conditions) increased in size through this second supply of amphiphiles, *i.e.*, the PANI-ES in the reaction mixture still maintained its "catalytic activity" for the incorporation of AOT molecules into the SUVs. The vesicle growth retaining a narrow size distribution, which is a specific growth behavior detected with PANI-ES, was again observed during the second supply of micelles (**Fig.4i**).

## Specific template interactions underlie selective growth

Although PANI-ES and PPy are cationic polymers while the supplied AOT and SDBS molecules are anionic (**Fig.1**), mere electrostatic attractions may not be sufficient to account for the observed vesicle growth, *i.e.*, a more specific "template" interaction may play a key role. To clarify this point, AOT vesicle growth experiments were carried out with the following three types of cationic polymers (**Fig.5a**): poly (diallyldimethylammonium chloride) (PDADMAC), poly-L-lysine hydrobromide (PLLHB), and poly (allylamine hydrochloride) (PAAHC). A freshly prepared 20 mM $NaH_2PO_4$ solution (pH = 4.3) containing 3.0 mM AOT (as SUVs) and approximately 1 mM monomer-equivalent cationic polymers was first equilibrated for 24 h at room temperature ($T$ ~ 25˚C). Then, micellar AOT solutions were supplied using the same protocol as in the "standard" vesicle growth experiments (**Fig.3a-h**), except that PANI-ES, PPy, and their precursor monomers were absent. **Figure 5b-d** show the resulting growth curves. In all three cases, vesicle growth was markedly weaker than that observed for the PANI- and PPy-containing variants. Specifically, the relative membrane area remained below 2.0 (*i.e.*, $\Delta a \lesssim +1.0$), comparable to that observed in the absence of polymeric additives (**Supplementary Figs.S1a, S2a,b**). By contrast, the corresponding PANI- or PPy-containing variants reached values of approximately 4.0 (*i.e.*, $\Delta a \sim +3.0$, **Fig.3a,e**). Therefore, PDADMAC, PLLHB, and PAAHC have almost no "catalytic" effect on AOT incorporation, in clear contrast to PANI-ES and PPy. These results indicate that the vesicle growth observed in **Fig.3a-h** cannot be explained solely by nonspecific electrostatic attraction, but rather requires specific interactions between the catalytic polymers and sulfonated amphiphiles.

## Discussion

Here, we discuss the growth behavior shown in **Fig.3a-h** from the perspective of compositional variation (or potentially evolution) features in a synthetic minimal cell system. In our previous system[18,19], the reproduction over 4–5 generations was achieved by supplying micellar AOT solutions and reactant molecules to a dispersion of target vesicles. The target vesicles were primarily composed of AOT and coupled with the enzyme-promoted chemical reaction network that synthesized PANI-ES on their vesicle surfaces. In the present study, we expanded this framework by systematically varying three components: template vesicle composition (AOT or SDBS+DA), catalytic polymer (PANI-ES or PPy), and supplied amphiphile (AOT or SDBS). These compositional variations generated diverse growth curves that are well described by the Hill equation (Eq. 1) with three parameters: $V_{max}$, $K_A$, $n$. This finding suggests that the synthetic minimal cell system with its reproduction ability can diversify and potentially evolve through compositional variations. In the following, we first discuss the physical basis of vesicle growth in terms of activation energy and chemical potential differences for incorporating supplied amphiphiles. We then consider how the Hill parameters may shape the fitness landscape of the synthetic minimal cell systems and thereby influence its evolutionary behavior.

The vesicle growth promotion and inhibition observed in **Fig.3a-h** are schematically summarized in **Fig.6a**, from which three notable features emerge.

(i) Both PANI-ES and PPy promote the incorporation of AOT into AOT or SDBS+DA(1/1) vesicles (a,d,e,h of **Figs.3, 6a**). Within the framework of the transition state theory, the rate of amphiphile incorporation into vesicle membranes under nonequilibrium steady state conditions can be expressed as

$$v_{\mathrm{D}} \sim v_0 e^{-\frac{u^{\ddagger}}{k_{\mathrm{B}}T}}\left(e^{\frac{\mu_{\mathrm{A}}}{k_{\mathrm{B}}T}} - e^{\frac{\mu_{\mathrm{V}}}{k_{\mathrm{B}}T}}\right) = v_0 e^{-\frac{\Delta u}{k_{\mathrm{B}}T}}\left(1 - e^{\frac{\Delta\mu}{k_{\mathrm{B}}T}}\right), \tag{2}$$

where $v_0$ is the attempt frequency for overcoming the activation barrier, $\Delta u = u^{\ddagger} - \mu_{\mathrm{A}}$ is the activation free-energy barrier for transferring amphiphiles from the bulk solution into the vesicle membrane (**Fig.6b**), $\Delta\mu = \mu_{\mathrm{V}} - \mu_{\mathrm{A}}$ is the chemical potential difference of amphiphiles in the bulk solution ($\mu_{\mathrm{A}}$) and vesicle membrane ($\mu_{\mathrm{V}}$), $k_{\mathrm{B}}$ is the Boltzmann constant, and $T$ is the temperature. In general, catalysts accelerate a process by lowering the activation free-energy barrier ($\Delta u$) without changing the overall thermodynamic driving force ($\Delta\mu$). In the present system, the pronounced growth promotion by PANI-ES or PPy, together with the suppressed growth with the reference cationic polymers (**Fig.5**) or without any polymeric additives (**Supplementary Figs.S1,S2**), suggest that specific template interactions between sulfonated amphiphiles and the catalytic polymers lower the effective activation barrier for amphiphile incorporation. As previously reported, if compared with typical hydrogen bonding, the hydrogen bonding between the negatively charged sulfonated head group ($-SO_3^-$) of AOT or SDBS and the positively charged amine radical cation of PANI-ES ($>N^{\cdot+}-H$) is considered to be significantly stronger due to additional electrostatic attraction[51,52]. The presence of such specific interaction appears the key factor for the observed efficient incorporation of the amphiphiles.

(ii) Both PANI-ES and PPy fail to promote the incorporation of SDBS into AOT vesicles (c,g of **Figs.3, 6a**). When SDBS was supplied as additional amphiphiles to the AOT(v)/PANI reaction mixtures (**Fig.3c**), the vesicle size remained nearly constant or exhibited a slight decrease. In contrast, in the AOT(v)/PPy reaction mixtures, increasing the concentration of supplied SDBS micelles resulted in a progressive decrease in vesicle size (**Fig.3g**). A similar decrease in vesicle size was also observed when the target mixture was simply an AOT SUV dispersion that did not contain the catalytic polymers or other additives (see the AOT(v)/SDBS(m) mixture in **Supplementary Fig.S1c**). These observations suggest that the supplied SDBS micelles do not promote vesicle growth, but instead induce partial extraction and solubilization of AOT molecules from the vesicle membrane into the bulk solution, possibly through the formation of mixed SDBS/AOT micellar aggregates. The more pronounced size decrease observed in the presence of PPy (**Fig.3g**) than PANI-ES (**Fig.3c**) further suggests that PPy facilitates this SDBS-induced solubilization process more effectively than PANI-ES, thereby enhancing the removal of AOT membrane components into the bulk solution.

(iii) PANI-ES promotes the incorporation of SDBS into SDBS+DA vesicles, whereas PPy does not, but instead induces vesicle shrinkage (b,f of **Figs.3, 6a**). In the absence of catalytic polymers, SDBS supply produced only limited growth ($\Delta a \sim +0.5$; **Supplementary Fig.S1b**), indicating a low net flux of SDBS into the membrane. In contrast, PANI-ES resulted in pronounced growth ($\Delta a \sim +4.0$; red bold line in **Fig.3b**), consistent with a reduction in the effective activation free-energy barrier for SDBS incorporation. PPy, however, caused a distinct decrease in vesicle size ($\Delta a \sim -0.7$; **Fig.3f**), suggesting that the combination of PPy and supplied SDBS favors the removal of SDBS or DA molecules from the membrane rather than their incorporation. As in enzyme kinetics, PANI-ES and PPy may enhance amphiphile flux by lowering the activation free-energy barrier ($\Delta u$). Unlike conventional enzymes, however, the direction of the net flux appears to depend on the polymer type. Because PANI-ES and PPy are localized at the vesicle surface through the template effect, they may modify the membrane chemical potential (*i.e.*, from $\mu_{\mathrm{V}}$ to $\mu_{\mathrm{V+P}}$ in Eq. 2) and thereby alter, or potentially reverse, the effective driving force between the vesicle membrane and the external dispersed state (*i.e.*, $\Delta\mu < 0$ or $\Delta\mu > 0$). Alternatively, the shrinkage observed with PPy may reflect enhanced SDBS-induced partial solubilization of membrane molecules, further biasing amphiphile flux toward the bulk solution.

This study explores a route toward evolvability in the synthetic minimal cell system developed in our previous studies[17,18], as a next stage beyond the previously demonstrated reproduction ability. Rather than demonstrating evolution itself, we focus here on whether a template-polymerization-based minimal cell system can generate multiple minimal cell variants with distinct growth properties. In this context, template vesicles and "catalytic" polymers (information polymers) form mutually catalytic relationships within a given environment and undergo vesicle reproduction, while compositional variations arise in both the vesicles and the polymers. As demonstrated in this study, the resulting minimal cell variants exhibited distinct growth curves (**Figs.3a-h, 6a**), providing differential, environment-dependent growth responses required as a basis for competitive selection.

An important point is that no single variant is universally superior because fitness is dynamically shaped by environmental conditions through the three composition-dependent Hill parameters. The magnitude of the maximal growth rate $V_{\max}$ may appear to serve as a metric for ranking the growth rates among the variants. However, this maximal rate is achieved only under sufficiently high concentrations of supplied amphiphiles, and that the saturation concentration itself varies among the variants. For example, the "SDBS+DA(v)/PANI/SDBS(m)" variant exhibited the highest maximal growth rate among all variants ($V_{\max}$ = 4.9 /10 min; **Table 1**); nevertheless, under conditions where the supplied amphiphile concentration was 2.0 mM, its growth rate was lower than that of the "AOT(v)/PANI/AOT(m)" variant (**Fig.3a,b**). In this context, the pseudo-dissociation constant $K_A$ becomes an important factor. This parameter governs how rapidly the sigmoidal growth curve begins to rise with respect to the supplied amphiphile concentration, and smaller values of $K_A$ allow the variants to achieve growth rates closer to $V_{\max}$ even at lower concentrations of amphiphiles. Therefore, a smaller $K_A$ is advantageous for competitive selection under "nutrient-poor" environments. Although the Hill coefficient $n$ does not directly reflect competitiveness to the same extent as the other two parameters, it refers to the steepness of the growth curve. Larger values of $n$ make the sigmoidal growth curve steeper, yielding a more pronounced threshold response. In other words, the system adopts a switch-like growth strategy; the variant remains in a stationary mode when the concentration of supplied amphiphiles is below a certain threshold value, but once this threshold value is exceeded, it rapidly transitions to growth at the maximal rate. Taken together, these growth behaviors indicate that the fitness of each synthetic minimal cell variant is better described in terms of fitness landscape rather than a simple one-dimensional hierarchy, with fitness dynamically shaped by the environmental conditions such as the availability of "food molecules" (supplied amphiphiles).

In conclusion, this study provides a physicochemical framework for establishing Darwinian evolution from simple molecular assemblies, by showing that the vesicle-polymer synthetic minimal cell variants can exhibit heritable-like molecular identity, differential growth, and environment-dependent selection potential. The next step is to allow these variants to compete directly within a simple bottom-up environment. For example, in a chamber initially dominated by AOT(v)/PANI/AOT(m) variant, continuous supply of AOT and aniline from one side of the chamber and SDBS, DA, and aniline from the other could generate spatially distinct minimal cell variants. If a newly generated variant exhibits a local growth advantage, it may increase in abundance and alter the population composition, thereby providing an experimentally accessible route toward evolutionary dynamics. Such a system would also enable direct investigation of how compositional inheritance changes during repeated growth and reproduction. In some cases, a variant may initially grow faster despite a mismatch between its vesicle composition and the supplied amphiphile. However, repeated growth and reproduction could gradually replace its vesicle components with food amphiphiles, thereby altering both, the vesicle's molecular identity and fitness. These coupled changes between compositional inheritance and competitive advantage highlights the potential of the present system for exploring dynamic, environmentally driven compositional evolution in synthetic minimal cells.

## Materials and Methods

### Materials

AOT (sodium bis-(2-ethylhexyl) sulfosuccinate, >99%, catalogue No. 86139) was purchased from Sigma-Aldrich Japan (Tokyo, Japan). SDBS (sodium dodecylbenzenesulfonate, hard type (mixture), >95%, No. D0990) was from Tokyo Chemical Industry (Tokyo, Japan). DA (decanoic acid, >99%, No. 041-23256) was from Wako Pure Chemical Industries (Osaka, Japan). The amphiphiles were used without further purification. SDBS and DA were dissolved in chloroform at 100 mM and stored at –20 °C as stock solutions.

Aniline (>99%), pyrrole (>99%), hydrogen peroxide ($H_2O_2$) (30% in water, ~9.8 M), sodium dihydrogenphosphate ($NaH_2PO_4$) dihydrate (>99.0%), phosphoric acid ($H_3PO_4$, >85%), acetone, acetonitrile, and chloroform (>99%) were purchased from Wako Pure Chemical Industries (Osaka, Japan). HRP (horseradish peroxidase, PEO-131, 286 U/mg, Lot No. 74590) was purchased from Toyobo Enzymes (Osaka, Japan). TvL (laccase from *Trametes versicolor*, Product No. 38429, 1.4 U/mg, Batch No. BCCK0654) and $ABTS^{2-}$ $(NH_4^+)_2$ (diammonium salt of 2,2'-azino-bis(3-ethylbenzothiazoline-6-sulfonate)) were from Sigma-Aldrich Japan (Tokyo, Japan). Cationic polymers poly-L-lysine hydrobromide (PLLHB) (MW = 1,000–5,000, No. P0879), poly (allylamine hydrochloride) (PAAHC) (MW ~ 17,500, No. 283215), and poly (diallyldimethylammonium chloride) (PDADMAC) (MW < 100,000, as aqueous solution, No. 522376) were from Sigma-Aldrich Japan (Tokyo, Japan). Ultrapure water purified with a Direct-Q 3 UV apparatus (Millipore, USA) was used to prepare all aqueous solutions and dispersions.

The concentration of HRP was determined spectrophotometrically using $\varepsilon_{403} = 1.02 \cdot 10^5\ M^{-1}cm^{-1}$ as molar absorption coefficient[53]. A TvL stock solution was prepared by centrifugating a 112 mg/mL TvL solution (prepared in 20 mM $NaH_2PO_4$, pH = 3.5) to remove insoluble materials. The supernatant solution was then used. The concentration of TvL was determined by measuring the initial enzymatic oxidation rate of $ABTS^{2-}$ in 20 mM $NaH_2PO_4$ solution (pH = 3.0) in the presence of dissolved $O_2$[28,54]. Briefly, various volumes of $ABTS^{2-}$ stock solutions (prepared in 20 mM $NaH_2PO_4$ solution (pH = 3.0)) were added to a defined volume of the TvL solution (prepared by diluting the centrifugated TvL solution with 20 mM $NaH_2PO_4$ solution (pH = 3.0)) to trigger the oxidation of $ABTS^{2-}$ to $ABTS^{\bullet-}$, quantified spectrophotometrically at 414 nm ($\varepsilon_{414}$ ($ABTS^{\bullet-}$) = $3.6 \cdot 10^4\ M^{-1}cm^{-1}$)[55]. The initial concentrations used were 5–160 μM $ABTS^{2-}$, 0.0167 mg/mL of TvL, and 20 mM of $NaH_2PO_4$ (pH = 3.0) at 25˚C. From the linear relationship between the initial increase in absorbance at 414 nm and TvL concentration, the molar concentration of the centrifuged 112 mg/mL TvL stock solution was estimated to 10 μM, and the $K_m$ value obtained for $ABTS^{2-}$ was 23 μM, which is comparable to the literature value of 60 μM, determined under similar conditions[56].

## Preparation of vesicles

AOT and binary SDBS+DA (1/1, molar ratio) small (sonicated) unilamellar vesicles (SUVs) with a radius of around 50 nm were used as template vesicles for the enzymatic polymerization of aniline and pyrrole. For the preparation of AOT SUVs, first, 17.8 mg of solid AOT was dissolved in 1 mL chloroform in a glass vial. Then, chloroform was evaporated by using a nitrogen gas stream under rotation of the vial by hand. A thin AOT film was formed on the inner glass surface of the vial. The AOT film was put under vacuum overnight to remove chloroform completely. The dried AOT film was hydrated by gently pouring 2.0 mL of a 20 mM $NaH_2PO_4$ solution into the vial (pH = 4.3 for the synthesis of PANI-ES and pH = 3.5 for the synthesis of PPy). The vial was then left in an incubation chamber at 60˚C for 1–2 h, followed by cooling down by keeping the vial at room temperature ($T$ ~ 25˚C). This procedure resulted in the formation of size-polydisperse AOT vesicles with radii of several micrometers (20 mM amphiphiles). It should be noted that in salt-free aqueous solution at this amphiphile concentration, AOT molecules do not form vesicles but form worm-like micelles[17]. The obtained AOT vesicle dispersion was subsequently homogenized with reduction in average vesicle size by applying ultrasonication using a Branson tip Sonifier model 150-D (Emerson, USA). Typically, three ten-second tip sonications (10 W power, 23 kHz frequency) were applied to a 2.0 mL AOT vesicle dispersion at room temperature (*T* ~ 25˚C) with one minute intervals to prevent excessive temperature increase. The hydrodynamic diameter of the AOT SUVs prepared with this procedure exhibited a rapid increase by several nanometers within a few hours after ultrasonication, reaching then an equilibrated size state. The obtained AOT SUVs had a mean hydrodynamic diameter of about 90 nm with a polydispersity of about 0.1. The vesicle growth experiments were carried out with such equilibrated dispersions, about 24 h after the preparation of the SUVs. Lamellarity, size-polydispersity, and spherical shape of AOT SUVs before use were analyzed by cryo-TEM (see below). Binary SDBS+DA (1/1) SUVs were prepared by the same procedure, except that the first SDBS+DA chloroform solution was not prepared by dissolving solid amphiphiles in chloroform but by mixing 200 μL of a 100 mM SDBS stock solution (in chloroform) with 200 μL of a 100 mM DA stock solution (in chloroform).

## Dynamic light scattering (DLS) measurements

Average size of AOT or SDBS+DA (1/1) vesicles were determined as hydrodynamic diameter by DLS measurements, assuming spherical vesicle shape. That the vesicles were spherical before and after the experiments was confirmed by cryo-TEM (see below). The DLS measurements were carried out at 90˚ scattering angle with an ALV-5000 goniometer system (ALV, Langen, Germany), an ALV-6000 multi-bit multi-tau correlator, and a diode-pumped laser Verdi V-2 (Coferent Inc., Santa Clara, USA) with vertically polarized light at $\lambda$ = 532 nm. The quartz cells used for the measurements were cleaned by dipping them in an ultrasonic bath that was filled with phosphorus-free detergent and ultrapure water, respectively. After drying in a heat bath at *T* ~ 40˚C, the quartz cells were cleaned in an ultrasonic bath with ethanol for 10 min and again dried. Finally, the cells were thoroughly washed with 0.2 μm filtered acetone and then dried. Before the measurements, the SUV dispersions were filtered with a 0.2 μm hydrophilic polytetrafluoroethylene (PTFE) Puradisc 25 filter (Cytiva, Tokyo, Japan) to remove possibly present dust and very large aggregates from the system. The typical sample volume for the DLS measurements was 1.5 mL. The intermediate scattering functions were recorded at fixed temperature *T* = 25˚C with a duration time of 30 s and data accumulation of 3–6 runs depending on the scattering intensities of the vesicle dispersions.

The time correlation function of the scattering intensity $I(t)$ is given by

$$\langle I(0)I(\tau)\rangle = \langle I\rangle^2 \left(1 + \left|g^{(1)}(\tau)\right|^2\right), \tag{3}$$

where $g^{(1)}(\tau)$ is the intermediate scattering function. $g^{(1)}(\tau)$ is described as

$$g^{(1)}(\tau) = \int_0^\infty e^{-\Gamma\tau} G(\Gamma) \mathrm{d}\Gamma \cong e^{-\bar{\Gamma}\tau}, \tag{4}$$

where $\Gamma$ and $\bar{\Gamma}$ are the initial relaxation rate and its mean value, respectively, and $G(\Gamma)$ is the distribution function of $\Gamma$. Assuming that the initial relaxation rate is governed by the diffusion of the mass center of the vesicles, the mean relaxation rate is express as $\bar{\Gamma} = \bar{D}q^2$, where $\bar{D}$ is mean diffusion constant, and $q = 2.23 \cdot 10^7\ \mathrm{m}^{-1}$ is the absolute value of the scattering vector in our measurements. Using the Stokes-Einstein equation for spherical vesicles, the mean diffusion constant is related to the mean hydrodynamic radius of vesicles $\bar{r}$ by

$$\bar{D} = \frac{k_\mathrm{B} T}{5\pi\eta\bar{r}}, \tag{5}$$

where $k_\mathrm{B}$ is the Boltzmann constant, $T$ is the temperature, and $\eta$ is the viscosity of the medium. Overall, the mean hydrodynamic radius $\bar{r}$ was obtained by measuring mean relaxation rates $\bar{\Gamma}$.

## Cryogenic transmission electron microscopy measurements

Lamellarity, size-polydispersity, and the shapes of SUVs were visualized in some of the reaction mixtures using cryo-TEM. Prepared AOT vesicle dispersions containing AOT SUVs (3.0 mM amphiphile) in 20 mM $NaH_2PO_4$ (pH = 4.3) were ultrasonicated and incubated for 24 h, corresponding to the state before the polymerization reaction. Then, a few μL aliquots of the sample solution were applied to a holey carbon film, Quantifoil R1.2/1.3 (Quantifoil Micro Tools, Germany) supported on a 200-mesh cupper grid. The grids were blotted with filter paper and frozen in liquid ethane using a semi-automatic EM GP2 plunger (Leica Microsystems, Germany) at 25°C and 95% relative humidity. Similarly, AOT SUVs/PANI-ES reaction mixtures were plunge-frozen for cryo-TEM observation 24 h after initiation of the polymerization reaction by the addition of HRP and $H_2O_2$, corresponding to the state prior to the vesicle growth experiments triggered by adding an AOT micellar solution. Finally, reaction mixtures containing AOT SUVs/PANI-ES prepared with HRP/$H_2O_2$ and supplemented with an AOT micellar solution were plunge-frozen 20 min after completion of the micelle addition, corresponding to the final state after AOT incorporation from the added AOT micelles.

The frozen grids were transferred to a CRYO ARM 300 II electron microscope (JEOL, Japan) equipped with a cold-field emission gun operated at 300 kV. The images were recorded with an in-column energy filter with a slit width of 20 eV and at a nominal magnification of × 25,000 on a Gatan K3 direct electron detector (AMETEK, U.S.A.). The nominal defocus range was 2 to 3.5 μm. Each image stack consisting of 50 dose-fractionated frames was exposed at a dose rate of 2.77 $e^{-}$ $Å^{-2}$ $sec^{-1}$ for 4 sec in CDS mode. The resultant frames were aligned and summed using DigitalMicrograph (AMETEK, U.S.A.).

## Vesicle surface-assisted synthesis of polyaniline and polypyrrole

Aniline or pyrrole were enzymatically polymerized in the presence of AOT or binary SDBS+DA (1/1) SUVs following a protocol described before[17,20,28]. For the polyaniline synthesis with HRP/$H_2O_2$, the following components of the reaction mixture were first added to 5.60 mL of a 20 mM $NaH_2PO_4$ solution (pH = 4.3) in a 10 mL glass vial at room temperature ($T$ ~ 25˚C): 1.20 mL of a SUV dispersion (20 mM amphiphiles in 20 mM $NaH_2PO_4$ solution, pH = 4.3), 0.80 mL of an aqueous aniline solution (40 mM in 20 mM $NaH_2PO_4$ solution, pH adjusted to 4.3 with $H_3PO_4$), and 0.40 mL of a HRP solution (18.4 μM in 20 mM $NaH_2PO_4$ solution, pH = 4.3). After gentle mixing, the polymerization reaction was triggered by quick addition of typically 9.00 μL of a freshly prepared $H_2O_2$ solution (2 M in water), again followed by gentle mixing. The initial reaction conditions were as follows: 3.0 mM amphiphiles (as SUVs), 4.0 mM aniline, 0.92 μM HRP, and typically 2.25 mM $H_2O_2$ in 20 mM $NaH_2PO_4$ (pH = 4.3). The total reaction time was 24 h. A homogeneous reaction mixture of dark green color was obtained. The optimal initial concentration of $H_2O_2$ is 4.5 mM for maximal aniline conversion (*i.e.*, polyaniline yield)[17,20]. However, the dense dark-green coloration of the reaction mixtures significantly attenuates the scattered light intensity, which hinders a reliable vesicle size analysis by DLS. Therefore, we reduced the $H_2O_2$ concentration typically by one half to lower the yield of polyaniline, which reduced coloration of the reaction mixtures but still ensured sufficient scattering intensity. The vesicle growth experiments and the DLS measurements were carried out 24 h after the addition of the $H_2O_2$ solution.

Polypyrrole was synthesized in the presence of template SUVs through the oxidation of pyrrole by the enzyme TvL and dissolved $O_2$ in 20 mM $NaH_2PO_4$ solution (pH = 3.5)[28]. First, 1.20 mL of an SUV dispersion (20 mM amphiphiles in 20 mM $NaH_2PO_4$ solution) and 0.80 mL of an aqueous pyrrole solution (40 mM in 20 mM $NaH_2PO_4$ solution, pH adjusted to 3.5 with $H_3PO_4$) were added to 4.52 mL of a 20 mM $NaH_2PO_4$ solution (pH = 3.5) in a 10 mL glass vial at room temperature ($T$ ~ 25˚C). After gentle mixing, the polymerization reaction was triggered by the addition of 1.48 mL of a TvL solution (supernatant solution after centrifugation; 10 μM in 20 mM $NaH_2PO_4$ solution), again followed by gentle mixing. The initial reaction conditions were as follows: 3.0 mM amphiphiles (as SUVs), 4.0 mM pyrrole, 1.85 μM TvL, and dissolved $O_2$ in 20 mM $NaH_2PO_4$ (pH = 3.5). The total reaction time was typically 24 h, and a homogeneous reaction mixture of green color was obtained. It should be noted that the rate of pyrrole polymerization with TvL/$O_2$ was much slower than the rate of aniline polymerization with HRP/$H_2O_2$; the pyrrole polymerization was continuing even after 24 h after initiating the reaction. The optimal reaction yield and the scattering intensity for performing the DLS measurements were adjusted by varying the reaction time, in contrast to the polyaniline synthesis experiments, where the reaction regulation occurred through a variation of the initial concentration of $H_2O_2$. In the case of the pyrrole/TvL system with dissolved $O_2$ as oxidant, the vesicle growth experiments and the DLS measurements were carried out typically 24 h after addition of the TvL solution.

## Characterization of polyaniline and polypyrrole

UV/Vis/NIR absorption measurements of the reaction mixtures were carried out with a V-730 spectrometer (JASCO, Japan) at $T$ ~ 25˚C, using quartz cuvettes with an optical path length of $L$ = 0.1 cm[17,20,28,57].

Raman absorption measurements of the reaction mixtures were carried out by using an inVia QONTOR confocal Raman spectrometer (Renishaw, UK), equipped with a diode-pumped solid-state laser ($\lambda$ = 532 nm, 50 mW), an optical microscope, and a CCD detector[18]. The polyaniline and polypyrrole reaction mixtures were dropped onto a borosilicate glass slide, and the Raman spectra were recorded in non-confocal mode with an objective lens (N Plan L50x, NA = 0.50 (Leica, Germany)). The exposure time for one measurement run was 1.0 s, and 100 runs were accumulated with about 15 mW laser power on the sample stage. All obtained spectra were fluorescence corrected with the WiRE software (Renishaw, UK).

## Quantification of remaining monomers

The amounts of remaining aniline or pyrrole in the reaction mixtures were determined by using the previously described methods[20,28,57]. The remaining amounts of aniline were quantified by withdrawing 30 µL of the reaction mixture and then adding it to 1470 µL acetonitrile into a 2 mL polypropylene microtube. After mixing and centrifugation, the UV/Vis/NIR absorption spectrum of the supernatant solution was measured using a quartz cuvette. The concentration of remaining aniline was calculated based on the absorption intensity at $\lambda$ = 238 nm and the molar absorption coefficient $\varepsilon_{238}$ (aniline) = $1.01 \cdot 10^4$ $M^{-1}cm^{-1}$. Here, the molar absorption coefficient value was determined from control experiments with known concentrations of aniline as described in Ref.[57]. The concentration of remaining pyrrole was similarly determined from the absorption intensity at $\lambda$ = 208 nm, using $\varepsilon_{208}$ (pyrrole) = $5.21 \cdot 10^3$ $M^{-1}cm^{-1}$, which was also determined from control experiments with known concentrations of pyrrole.

## Analysis of vesicle growth in the presence of enzymatically formed polyaniline or polypyrrole

AOT or SDBS micellar solutions were supplied to SUV dispersions containing enzymatically formed polyaniline or polypyrrole in 20 mM $NaH_2PO_4$ solution (**Fig.3a-h**). The AOT micellar solutions (3–20 mM AOT) were prepared by simply dissolving solid AOT in ultrapure water, followed by mixing with the help of a vortex mixer. It should be noted that under the conditions used, AOT molecules do not form vesicles in water in the absence of $NaH_2PO_4$, but assemble into micelles, which are in equilibrium with non-assembled AOT molecules. SDBS micellar solutions (3–20 mM SDBS) were prepared similarly in ultrapure water. All micellar solutions were filtrated before use with a 0.2 μm hydrophilic PTFE filter. For the vesicle growth experiments, first, 1.0 mL of the PANI-ES or PPy reaction mixture containing SUVs was dispensed into a 5 mL microtube after 24 h from initiating the reaction. Before supplying the micellar solutions, the reaction mixtures contained 3.0 mM amphiphiles (as SUVs), enzymatically formed PANI-ES or PPy, remaining monomers (aniline or pyrrole), and enzymes (HRP or TvL) in 20 mM $NaH_2PO_4$ (pH = 4.3 or 3.5). Then, 100 μL of a micellar solution (3–20 mM amphiphile) and 100 μL of a 40 mM $NaH_2PO_4$ solution (pH = 4.3 or 3.5) were quickly added by using a micropipette. After 30 s, the procedure was repeated four times with intervals of 30 s. Overall, 500 μL of a micellar solution and 500 μL of a 40 mM $NaH_2PO_4$ solution were supplied within two minutes. This resulted in a doubling of the total volume of the reaction mixtures to 2.0 mL. During the supply of amphiphiles, the reaction mixtures were gently stirred at room temperature (*T* ~ 25˚C) using a magnetic stirrer with a micro-rotor. The concentration of $NaH_2PO_4$ in the reaction mixtures was kept constant during the experiments. This is considered important because a change of the total ion concentration may influence the self-assembly property of the amphiphiles, *e.g.*, critical aggregation concentrations[17]. At 10 min after completing the addition of micelles, the mean size of the vesicles present in the reaction mixtures was determined by DLS measurements as hydrodynamic vesicle radius. Vesicle growth was estimated by determining the increase of the relative membrane area $a$:

$$a(A) = 1 + \Delta a(A) = \frac{S_0 + \{S(A) - S_0\}}{S_0} = \frac{R(A)^2}{R_0^2}, \tag{6}$$

where $A$ is the concentration of the supplied amphiphiles in the reaction mixtures, $\Delta a$ is the relative vesicle growth rate, $R_0$ and $R$ are the mean hydrodynamic radii measured before and 10 min after supplying additional amphiphiles, respectively, and $S_0$ and $S$ are the corresponding membrane areas calculated from the determined mean hydrodynamic radii, assuming spherical vesicle shapes.

In three control experiments, the micellar solutions and the 40 mM $NaH_2PO_4$ solutions (pH = 4.3 or 3.5) were similarly added to (i) SUV dispersions (3.0 mM amphiphiles as SUVs in 20 mM $NaH_2PO_4$ solution) without any further compounds added (**Supplementary**

**Fig.S1**); (ii) SUV dispersions containing aniline or pyrrole (3.0 mM amphiphiles and 4.0 mM aniline or pyrrole in 20 mM $NaH_2PO_4$ but no polymerization reaction taking place; **Supplementary Fig.S2**); and (iii) SUV dispersions containing other cationic polymers (3.0 mM amphiphiles and 1 mM monomer-equivalent concentration of cationic polymers in 20 mM $NaH_2PO_4$; **Fig.5**).

**Data availability**

The datasets generated and analyzed during the present study are available in the Kyushu University Institutional Repository (QIR), doi: XXX.

**Acknowledgements**

This research was supported by JSPS KAKENHI (grant number JP22K20346, JP23H00087, and JP23K13070), JST ACT-X (grant number JPMJAX23DA), and Research Support Project for Life Science and Drug Discovery (Basis for Supporting Innovative Drug Discovery and Life Science Research (BINDS)) from AMED (grant Number JP25ama121006).

**Author contributions**

M.K., P.W., and M.I. conceived the work. M.K. and T.S. synthesized PANI-ES and PPy. M.K. and P.W. characterized PANI-ES and PPy. M.K., T.S., R.K., and K.M. conducted the vesicle growth experiments and the DLS measurements. M.K., D.U., T.H., and K.Y. carried out the cryo-TEM measurements. M.K. and T.S. analyzed the experimental data. M.K. wrote the original manuscript. M.K., P.W., and M.I. wrote the final manuscript with input from all authors.

**Declaration of competing interest**

The authors declare that they have no known competing financial interests or personal relationships that could have influenced the work reported in this paper.

## Figures

**Figure 1:**

**a**. Chemical structures of the amphiphiles used to form vesicles: sodium bis-(2-ethylhexyl) sulfosuccinate (AOT), sodium dodecylbenzenesulfonate (SDBS), and decanoic acid (DA). The branched hydrophobic tail of SDBS is a representative structure of the different isomers present in commercial SDBS[29].

**b**. Chemical structures of the catalytic polymers synthesized enzymatically: polyaniline in its emeraldine salt form (PANI-ES) and polypyrrole (PPy). The structures shown are the bipolaron forms[20,28]. $A^{\ominus}$ represents a counter ion, the sulfonated amphiphiles shown in **a**.

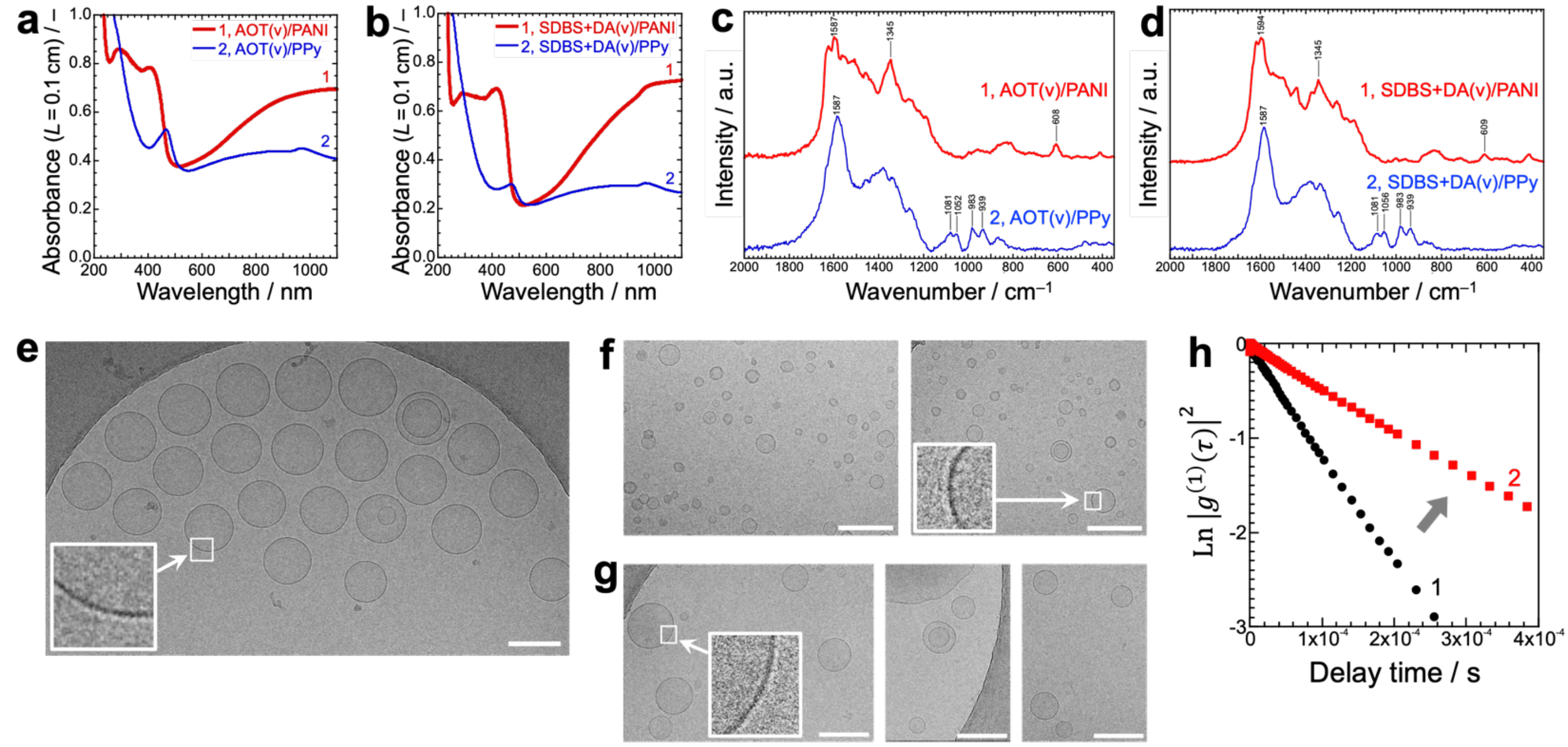

**Figure 2: Characterization of the obtained PANI-ES and PPy, and measurement of vesicle shape and size.**

**a**. UV/Vis/NIR absorption spectra of reaction mixtures containing polyaniline (PANI-ES, red bold lines, 1) or polypyrrole (blue thin lines, 2) that were enzymatically synthesized in the presence of AOT SUVs, and

**b**. binary SDBS+DA (1/1) SUVs, recorded at *t* = 24 h after initiation of the reactions. The abbreviations used for the samples analyzed are "AOT(v)/PANI", "AOT(v)/PPy", "SDBS+DA(v)/PANI", and "SDBS+DA(v)/PPy", with "(v)" indicating that the templates used were vesicles (SUVs).

**c**. Raman spectra of reaction mixtures containing polyaniline (PANI-ES, red bold lines, 1) or polypyrrole (blue thin lines, 2) that were enzymatically synthesized in the presence of AOT SUVs, and

**d**. binary SDBS+DA (1/1) SUVs, recorded at *t* = 24 h after initiation of the reactions.

**e**. Typical cryo-TEM images of AOT SUVs dispersion before the reaction ([AOT] = 3.0 mM in 20 mM $NaH_2PO_4$ solution (pH = 4.3)). Plunge-frozen 24 h after preparation by ultrasonication. Scale bar: 100 nm.

**f**. Typical cryo-TEM images of AOT SUVs containing enzymatically synthesized PANI-ES (sample "AOT(v)/PANI") *before* the vesicle growth experiments were initiated. Plunge-frozen 24 h after starting the enzymatic polymerization reaction. Scale bar: 100 nm.

**g**. Typical cryo-TEM images of AOT SUVs containing enzymatically synthesized PANI-ES *after* the vesicle growth experiment (sample "AOT(v)/PANI/AOT(m)"). Frozen 20 min after completion of the micelle addition ([supplied AOT] = +3.75 mM). Scale bar: 100 nm.

**h**. Examples of the intermediate scattering functions of a dispersion of SUVs containing enzymatically synthesized PANI-ES, analyzed by DLS 24 h after initiating the aniline polymerization reaction (sample “AOT(v)/PANI”, black filled circles, 1) and after supplying micellar AOT solutions (sample "AOT(v)/PANI/AOT(m)”, red filled squares, 2). The two relaxation rates correspond to mean hydrodynamic diameters of 39.2 nm and 78.8 nm, respectively. The gray arrow indicates the shift of the relaxation rate, reflecting vesicle growth.

For **a**–**d**,**f**–**h**, the initial reaction conditions for the synthesis of PANI-ES were [amphiphile] = 3.0 mM, $[\text{aniline}]_0$ = 4.0 mM, [HRP] = 0.92 μM, and $[H_2O_2]_0$ = 2.25 mM in 20 mM $NaH_2PO_4$ solution (pH = 4.3). The initial reaction conditions for the synthesis of PPy were [amphiphile] = 3.0 mM, $[\text{pyrrole}]_0$ = 4.0 mM, [TvL] = 1.85 μM, and dissolved oxygen in 20 mM $NaH_2PO_4$ solution (pH = 3.5). All reactions were carried out at room temperature ($T$ ~ 25˚C), and the measurements were carried out after $t$ = 24 h from initiating the reactions.

For **e**–**g**, insets show 5x enlarged images of the region indicated by arrows. The defocus settings for the cryo-TEM images were **e**. –3 μm, **f**. –3.5 μm, and **g**. –2 μm.

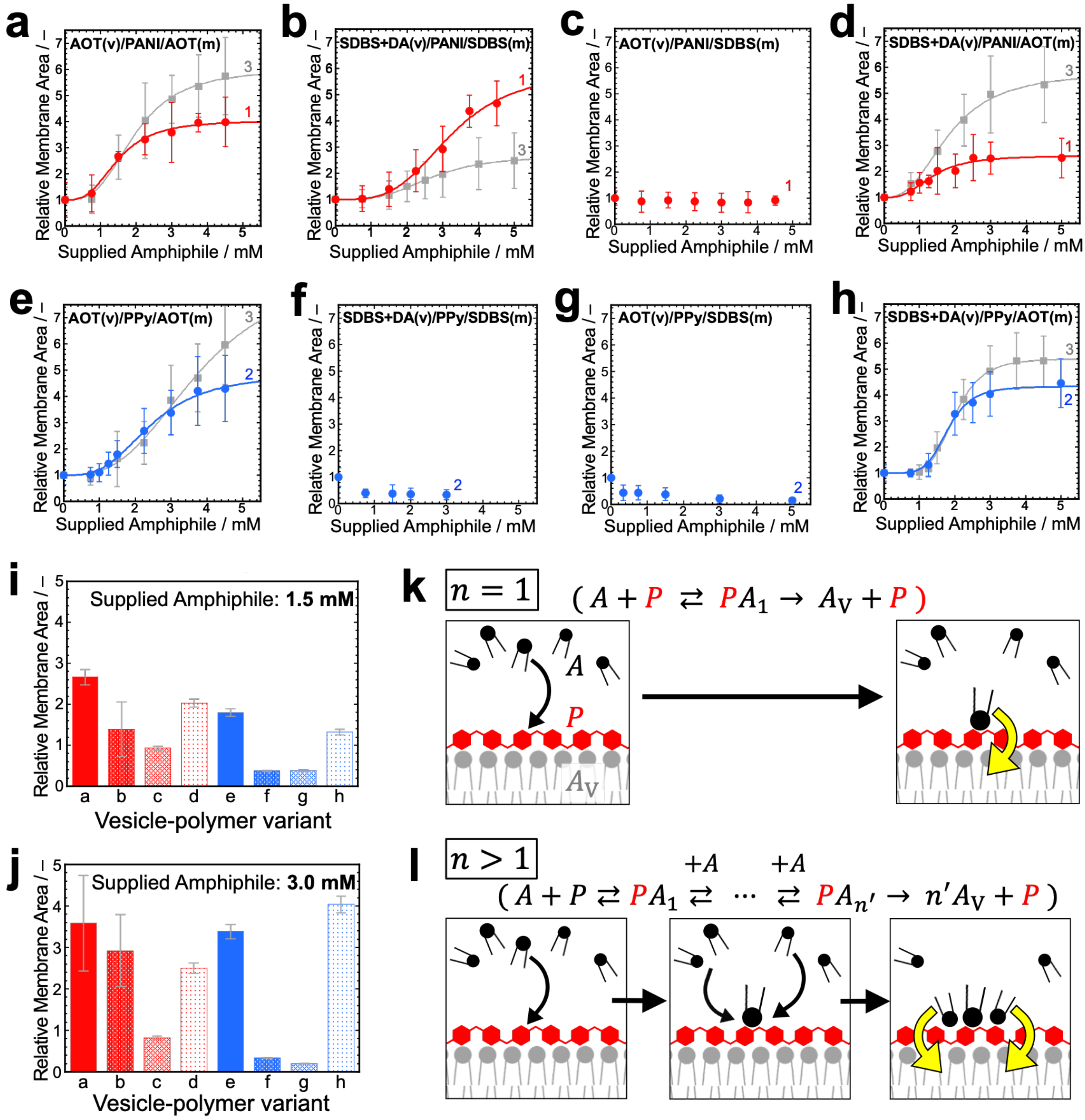

**Figure 3: Growth curves of synthetic minimal cell variants.**

Relative membrane area of AOT SUVs or binary SDBS+DA (1/1) SUVs in response to the addition of AOT or SDBS micellar solutions to the corresponding reaction mixtures containing enzymatically synthesized PANI-ES (red filled circles, 1) or PPy (blue filled circles, 2). The gray filled squares (3) are data that were obtained under conditions that deviated from the typical conditions, with either higher amounts (a,d,e,h) or lower amounts (b) of enzymatically formed PANI-ES or PPy (see main text).

**a**. AOT SUVs with PANI-ES and the supply of micellar AOT, abbreviated as "AOT(v)/PANI/AOT(m)";

**b**. SDBS+DA (1/1) SUVs with PANI-ES and the supply of micellar SDBS, abbreviated as "SDBS+DA(v)/PANI/SDBS(m)";

**c**. AOT SUVs with PANI-ES and the supply of micellar SDBS;

**d**. SDBS+DA (1/1) SUVs with PANI-ES and the supply of micellar AOT;

**e**. AOT SUVs with PPy and the supply of micellar AOT;

**f**. SDBS+DA (1/1) SUVs with PPy and the supply of micellar SDBS;

**g**. AOT SUVs with PPy and the supply of micellar SDBS;

**h**. SDBS+DA (1/1) SUVs with PPy and the supply of micellar AOT.

**i**. Comparison of relative membrane area for the eight experimental conditions shown in **a–h** at the supplied amphiphile concentrations of 1.5 mM, and

**j**. at the supplied amphiphile concentrations of 3.0 mM.

**k**. Suggested reaction scheme for the vesicle growth with non-cooperative incorporation, and

**l**. cooperative incorporation of the supplied amphiphiles (black "$A$") into the vesicle membrane (gray "$A_{\mathrm{V}}$") *via* binding to PANI-ES or PPy as "catalytic polymer" (red "$P$"). "$PA_i$" represents an intermediate complex in which $i$ amphiphiles are clustered on the catalytic polymer. It should be noted that a Hill coefficient value $n(>1)$ does not directly reflect the actual number of amphiphiles clustering on the catalytic polymers $n'$, but rather represents the degree of cooperativity[49,50]. The curved black arrows in the drawings represent the binding of amphiphiles to PANI-ES or PPy, and the yellow arrows represent the incorporation of amphiphiles into the vesicle membranes.

For **a**–**h**, the horizontal axis represents the concentration of supplied amphiphiles in the final reaction mixtures. The initial mean hydrodynamic diameters of the SUVs before supplying additional amphiphiles were approximately 45 ± 10 nm. The DLS data were recorded 10 min after completion of the supply of amphiphile micelles. The error bars represent standard deviations obtained from three (for vesicle growth) or two (for no growth) observations. The sigmoidal vesicle growth was fitted by the Hill equation (Eq. 1), and the fitting parameter values are listed in **Table 1**.

**Table 1:**

The fitting parameter values obtained from the data shown in **Fig.3a–h** by employing the Hill equation (Eq. 1). The values obtained for the typical conditions of the enzymatically synthesized PANI-ES or PPy (red and blue circles in **Fig.3a–h**) are listed; the values in the brackets were obtained under conditions that deviated from the typical conditions (gray squares in **Fig.3a–h**). "AOT(v)/PANI/AOT(m)" stands for a vesicle growth experiment in which AOT SUVs were used as templates for the polymerization of aniline with HRP and $H_2O_2$ to PANI-ES, followed by addition of a micellar AOT solution. "SDBS+DA(v)/PPy/SDBS(m)" stands for a vesicle growth experiment in which SDBS+DA(1/1, molar ratio) SUVs were used as templates for the polymerization of pyrrole with TvL and dissolved $O_2$ to PPy, followed by addition of a micellar SDBS solution.

| Fig. 6 | Composition | $\boldsymbol{n}$ [−] | $\boldsymbol{K_A}$ [mM] | $\boldsymbol{V_{\max}}$ [/10 min] |
|---|---|---|---|---|
| **a** | AOT(v) / PANI / AOT(m) | 3.0 (3.1) | 1.5 (2.0) | 3.0 (5.0) |
| **b** | SDBS-DA(v) / PANI / SDBS(m) | 3.6 (3.4) | 3.2 (2.7) | 4.9 (1.7) |
| **c** | AOT(v) / PANI / SDBS(m) | – | – | – |
| **d** | SDBS-DA(v) / PANI / AOT(m) | 2.9 (2.6) | 1.3 (1.8) | 1.6 (4.8) |
| **e** | AOT(v) / PPy / AOT(m) | 3.1 (2.9) | 2.5 (3.8) | 3.9 (7.9) |
| **f** | SDBS-DA(v) / PPy / SDBS(m) | – | – | – |
| **g** | AOT(v) / PPy / SDBS(m) | – | – | – |
| **h** | SDBS-DA(v) / PPy / AOT(m) | 5.3 (5.2) | 1.8 (2.0) | 3.3 (4.4) |

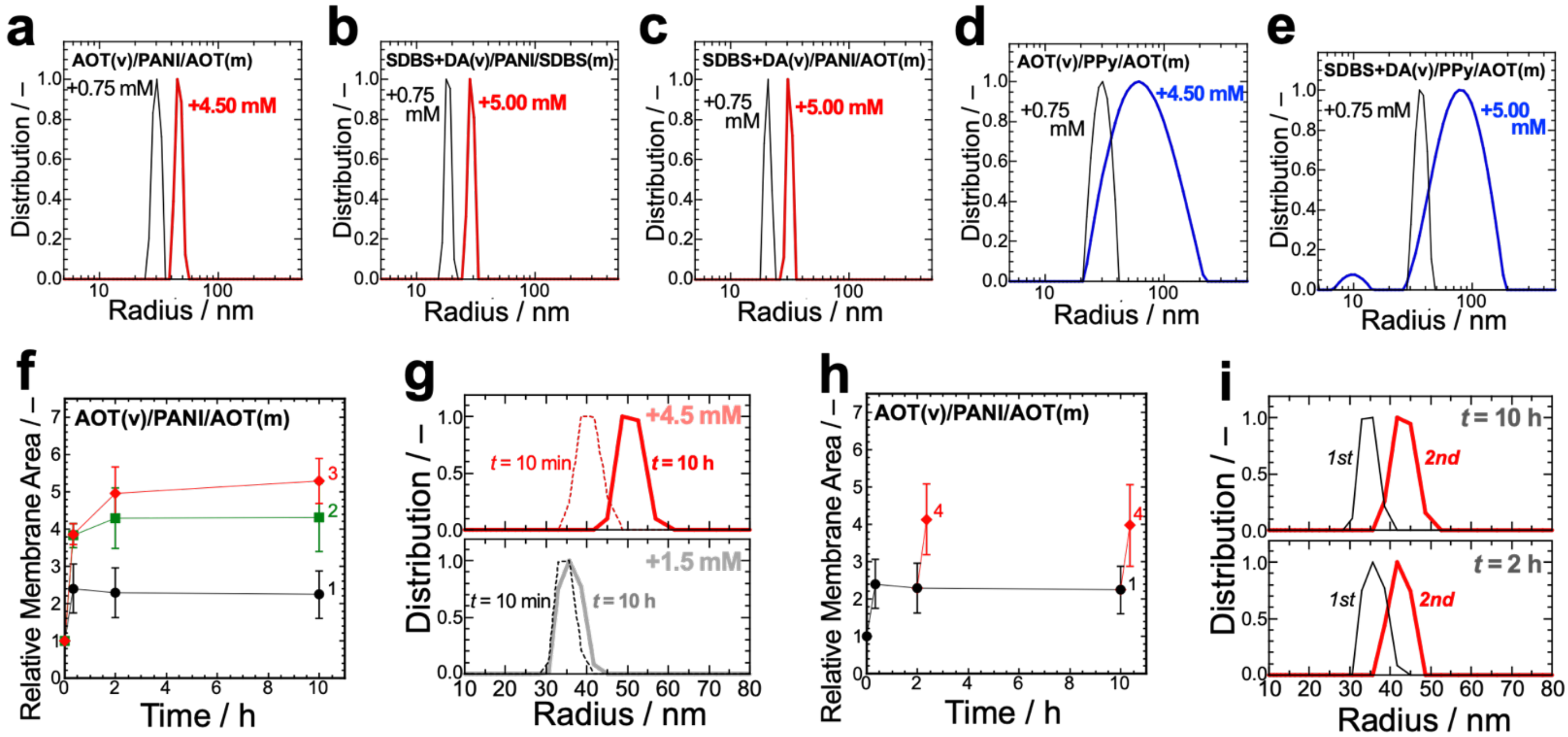

**Figure 4: Vesicle size distributions and growth kinetics analysis.**

**a**. CONTIN plots (*i.e.*, distribution of hydrodynamic radii) obtained from the intermediate scattering functions used for drawing **Fig.3a–h**. AOT SUVs with enzymatically synthesized PANI-ES and the supply of micellar AOT ("AOT(v)/PANI/AOT(m)");

**b**. SDBS+DA (1/1) SUVs with enzymatically synthesized PANI-ES and the supply of micellar SDBS;

**c**. SDBS+DA (1/1) SUVs with enzymatically synthesized PANI-ES and the supply of micellar AOT;

**d**. AOT SUVs with enzymatically synthesized PPy and the supply of micellar AOT;

**e**. SDBS+DA (1/1) SUVs with enzymatically synthesized PPy and the supply of micellar AOT.

In each figure, vesicle size distributions are compared for two concentrations of supplied amphiphiles, a lower one (thin black lines, +0.75 mM) and a higher one (bold red lines for PANI-ES and bold blue lines for PPy, +4.50 or +5.00 mM).

**f**. Time dependence of the relative surface area of AOT SUVs containing enzymatically formed PANI-ES after supplying AOT micelles. The mean hydrodynamic radii of the SUVs were determined 10 min, 2 h, and 10 h after the addition of micellar AOT solutions. The concentrations of the supplied AOT in the reaction mixtures were 1.5 mM (black filled circles, 1), 3.0 mM (green filled squares, 2), and 4.5 mM (red filled diamonds, 3).

**g**. CONTIN plots (*i.e.*, size distributions) obtained from the intermediate scattering functions for drawing **f**. The upper and lower panels show the size distributions when +4.5 mM and +1.5 mM additional AOT was supplied, respectively. The size distributions were determined 10 min (thin dotted lines) and 10 h (bold solid lines) after completion of the micelle supply.

**h**. To portions of a reaction mixture experiencing the first supply of 1.5 mM AOT, additional 1.5 mM of AOT micelles were supplied 2 h and 10 h after the first supply was completed (red filled diamonds, 4). The black filled circles (1) are the same data as in **f**.

**i.** CONTIN plots (*i.e.*, size distributions) obtained from the intermediate scattering functions for drawing **h**. The upper and lower panels show the size distribution determined at 10 h and 2 h, respectively, by withdrawing portions of the reaction mixtures that already experienced micelle addition (+1.5 mM). The thin lines and the bold lines represent the size distributions before (“1st”) and 10 min after the second supply (“2nd”) of +1.5 mM amphiphiles.

In **e**, the supply of AOT molecules into the SDBS+DA(v)/PPy mixtures consistently and reproducibly led to a small peak at $r$ ~ 10 nm, implying that the supplied AOT micelles either remained unincorporated or assembled into new aggregates distinct from the pre-existing SUVs.

For **f** and **h**, error bars represent standard deviations obtained from three observations. The lines connecting the data points are guides to the eye only.

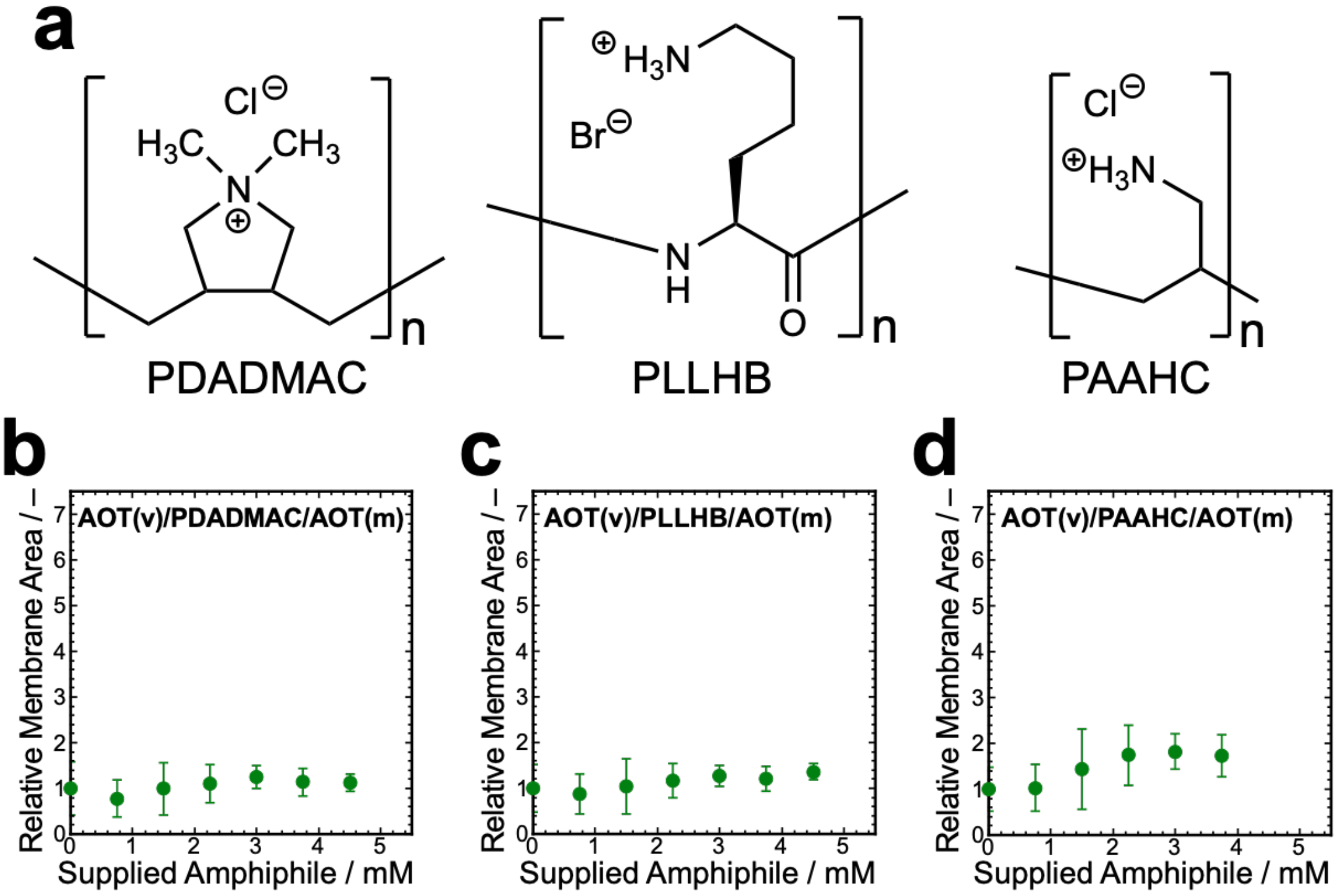

**Figure 5: Control growth experiments with nonspecific cationic polymers.**

**a.** Chemical structures of the cationic polymers used in the control measurements; poly (diallyldimethylammonium chloride) (PDADMAC), poly-L-lysine hydrobromide (PLLHB), and poly (allylamine hydrochloride) (PAAHC). They all carry positive charges at pH = 4.3 (in 20 mM $NaH_2PO_4$).

**b**. Values of the relative membrane area of AOT SUVs in the presence of PDADMAC,

**c**. PLLHB, and

**d**. PAAHC after supplying AOT micelles.

For **b–d**, the target SUV dispersion contained 3.0 mM AOT and approximately 1 mM monomer-equivalent concentrations of cationic polymers in 20 mM $NaH_2PO_4$ solution (pH = 4.3). The DLS data were recorded 10 min after completion of the supply of AOT micelles. Error bars represent standard deviations obtained from three observations. ”AOT(v)/PDADMAC/AOT(m)” stands for “AOT vesicle dispersion to which the cationic polymer PDADMAC and a micellar AOT solution were added”.

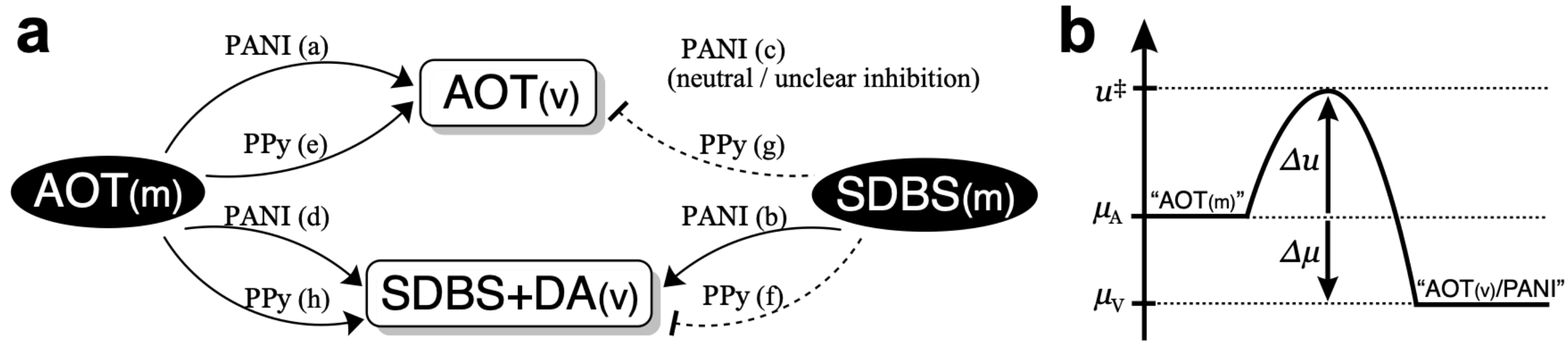

**Figure 6: Schematic summary of the observed vesicle growth promotion or inhibition.**

**a**. The supply of micellar AOT solution is abbreviated as "AOT(m)", and the template AOT vesicle is abbreviated as "AOT(v)". The solid curved arrows represent the growth of the targeted vesicles being promoted in the presence of PANI-ES or PPy on the vesicle surfaces (head of the arrow), by supplying the micelles (tail of the arrow). The dashed bar-headed arrows similarly represent the inhibition of vesicle growth (*i.e.*, size decrease) of the targeted vesicles. The labels (a) to (h) refer to the variants and experimental data of **Fig.3a-h**.

**b**. Schematic drawing of the activation free energy barrier ($\Delta u$) and the chemical potential difference ($\Delta\mu$) for the supplied amphiphiles between the initial state and the final state, with the chemical potentials $\mu_{\mathrm{A}}$ and $\mu_{\mathrm{V}}$, respectively (Eq. 2). $u^{\ddagger}$ represents the transition state energy. The case of vesicle growth with the "AOT(v)/PANI/AOT(m)" composition (**Fig. 3a**) is shown representatively, where $\Delta\mu < 0$ provides a thermodynamic driving force for incorporating the supplied AOT molecules into the vesicle membrane. The catalytic polymer PANI-ES reduces the activation barrier, resulting in an enhanced net forward flux (*i.e.*, enhanced vesicle growth).

# Supplementary Information

## Vesicle-surface-templated catalytic polymers drive differential growth in synthetic minimal cell variants.

Minoru Kurisu[a,b,*], Taro Suzuki[a], Ryosuke Katayama[a], Kazuki Maruyama[a], Daisuke Unabara[c], Tasuku Hamaguchi[c], Koji Yonekura[c,d], Peter Walde[e], and Masayuki Imai[a]

[a] Department of Physics, Graduate School of Science, Tohoku University, 6-3 Aramaki, Aoba, Sendai 980-8578, Japan.

[b] Department of Physics, Faculty of Science, Kyushu University, 744 Motooka, Nishi-ku, Fukuoka 819-0395, Japan.

[c] Institute of Multidisciplinary Research for Advanced Materials, Tohoku University, 2-1-1 Katahira, Aoba-ku, Sendai 980-8577, Japan.

[d] RIKEN SPring-8 Center, 1-1-1 Kouto, Sayo, Hyogo 679-5148, Japan.

[e] Department of Materials, ETH Zürich, Leopold Ruzicka-Weg 4, CH-8093 Zürich, Switzerland.

*Corresponding author. E-mail: kurisu.minoru.877@m.kyushu-u.ac.jp

## Supplementary Note1:

## Control experiments for better understanding the observed vesicle growth

The observed growth of template SUVs in the reaction mixtures in the presence of PANI-ES or PPy (*e.g.*, “AOT(v)/PANI/AOT(m)” variant; **Fig.3a**-**h**) was compared with the results obtained in control experiments lacking PANI-ES or PPy but still involving the supply of sulfonated amphiphiles (*e.g.*, control experiments using “AOT(v)/AOT(m)”).

### (i) Control 1: No PANI or PPy synthesis in vesicle dispersions

First, micellar AOT or SDBS solutions were supplied to dispersions that contained only AOT SUVs or binary SDBS+DA (1/1) SUVs in 20 mM $NaH_2PO_4$ solution (*i.e.*, the template SUV dispersions as prepared). The procedure of supplying the amphiphiles was the same as in the case of the growth observations of **Fig.3a**-**h** except that the target dispersions did not contain the reaction components: 3.0 mM amphiphiles (AOT or SDBS+DA (1/1)) as SUVs in 20 mM $NaH_2PO_4$ solution (pH = 4.3 or 3.5). 500 μL in total of micellar AOT or SDBS solutions (3–20 mM amphiphile) and 500 μL in total of a 40 mM $NaH_2PO_4$ solution were added within two minutes under gentle mixing into 1.0 mL of the target vesicle dispersions. The results are shown in **Fig. S1a–d**. In the absence of PANI-ES or PPy, the supply of additional sulfonated amphiphiles to AOT or SDBS+DA (1/1) SUVs resulted in only limited vesicle growth compared to the data shown in **Fig.3a**-**h**, indicating that the amphiphile addition alone is insufficient to account for the observed growth in the presence of PANI-ES or PPy.

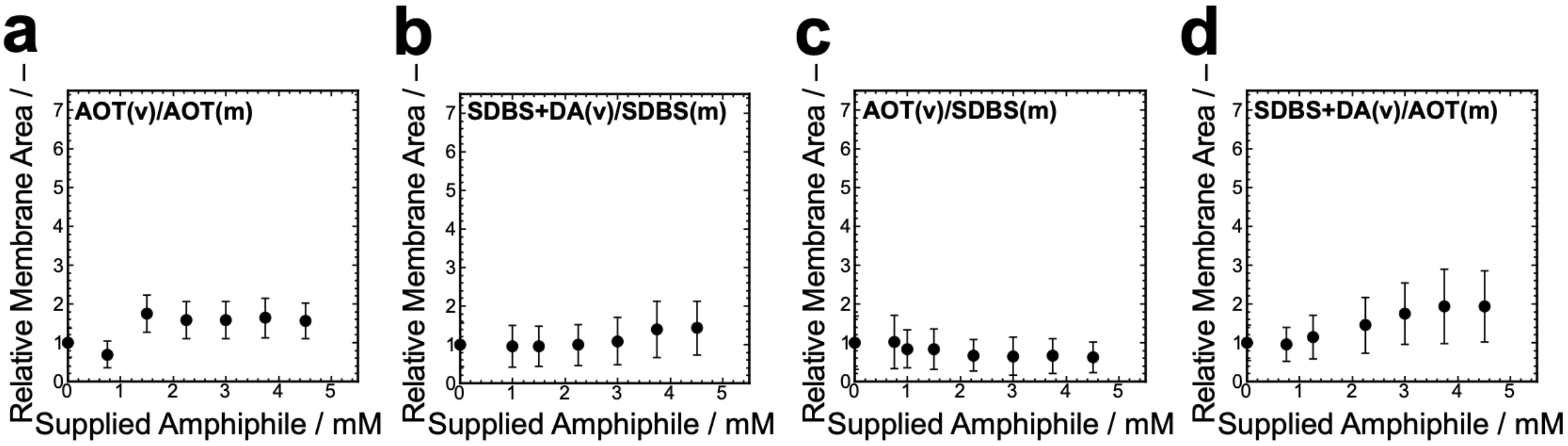

**Figure S1:**

Relative membrane area of AOT SUVs or binary SDBS+DA (1/1) SUVs upon the addition of micellar AOT or SDBS solutions in the absence of any reaction components (*i.e.*, without formation of PANI-ES or PPy). **a**. AOT SUVs with the supply of micellar AOT (“AOT(v)/AOT(m)”); **b**. SDBS+DA (1/1) SUVs with the supply of micellar SDBS; **c**. AOT SUVs with the supply of micellar SDBS; **d**. SDBS+DA (1/1) SUVs with the supply of micellar AOT. The DLS data were recorded 10 min after completion of the supply of the micellar solutions. Error bars represent standard deviations obtained from three independent measurements.

**(ii) Control 2: No PANI or PPy synthesis, but aniline or pyrrole present in vesicle dispersions**

Another presumed factor that may promote vesicle growth, aside from the presence of PANI-ES or PPy on the vesicle surface, is the remaining amounts of monomers (aniline or pyrrole) in the reaction mixtures. In the vesicle growth observations described in the main text, we either decreased (i) the amount of reaction trigger ($H_2O_2$) for the PANI-ES synthesis or (ii) the reaction time for the PPy synthesis from the optimal values obtaining maximal yield of PANI-ES or PPy, thereby suppressing the reaction yield and reducing the coloration of the reaction mixture. These adjustments ensured sufficient scattering intensity for the DLS measurements, which also led to an increase in the amounts of unreacted monomers. The reaction mixtures initially contained 4.0 mM aniline or pyrrole at the beginning of the reactions, and the quantification of the remaining monomers indicated that approximately half of the monomers in the reaction mixtures remained unreacted during the vesicle growth observations reported in the main text (**Fig.3a**-**h**). The physical properties of the template SUVs such as membrane fluidity, surface polarity, and membrane phase state (liquid-ordered or liquid-disordered) can be influenced by these remaining monomers[S1], which might also influence the vesicle growth rate. Therefore, a control experiment for identifying the effect of remaining monomers was carried out: AOT or SDBS micellar solutions were added to vesicle dispersions containing 3.0 mM amphiphiles (as SUVs) and 4.0 mM aniline or pyrrole in 20 mM $NaH_2PO_4$ solution (pH = 4.3 or 3.5). The only difference with respect to the growth observations (**Fig.3a**-**h**) was the composition of the target vesicle dispersion. The results are shown in **Fig. S2a**-**d**. The relative membrane area remained at around 1.0 and did not exceed 2.0. These results indicate that the presence of monomeric aniline and pyrrole in the reaction mixtures cannot be a dominating factor to induce the remarkable vesicle growth shown in **Fig.3a**-**h**.

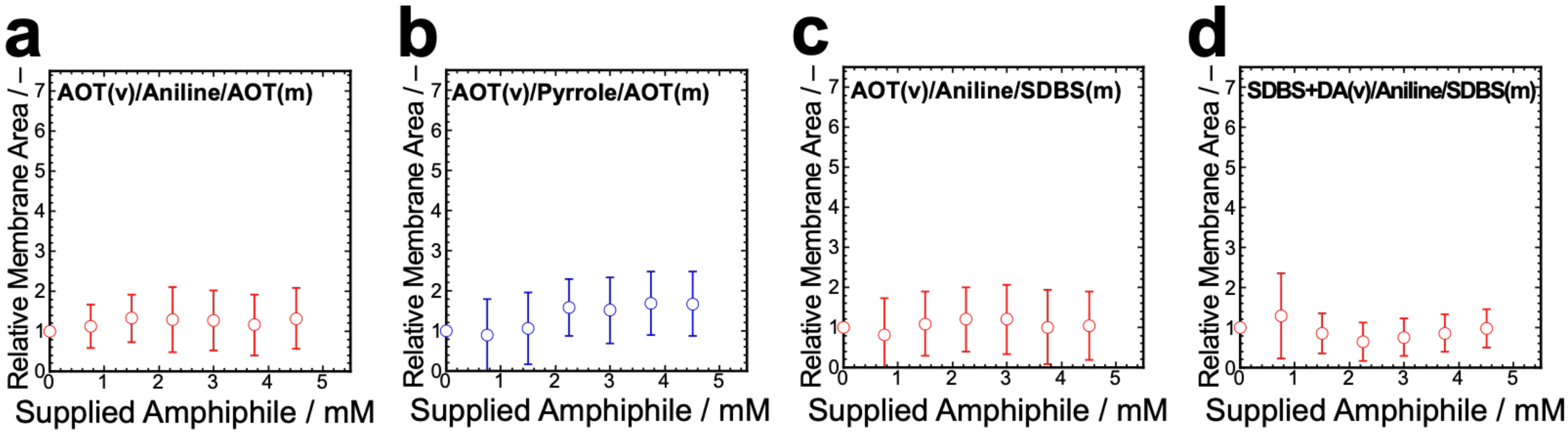

**Figure S2:**

Relative membrane area of AOT SUVs or binary SDBS+DA (1/1) SUVs containing 4.0 mM aniline or pyrrole, upon addition of micellar AOT or SDBS solutions. **a**. addition of AOT micelles to AOT SUVs containing aniline ("AOT(v)/Aniline/AOT(m)"); **b**. addition of AOT micelles to AOT SUVs containing pyrrole; **c**. addition of SDBS micelles to AOT SUVs containing aniline; **d**. addition of SDBS micelles to SDBS+DA (1/1) SUVs containing aniline. The DLS data were recorded 10 min after completion of the supply of the micellar solution. Error bars represent standard deviations obtained from three independent measurements.